%% file: main.tex
\pdfoutput=1
\documentclass[letterpaper, 10 pt, conference]{ieeeconf}

\usepackage{times}
\usepackage{color, colortbl}
\usepackage[dvipsnames]{xcolor}
\usepackage[bookmarks=true]{hyperref}
\usepackage{amsfonts}
\usepackage{amsmath}
\usepackage{amssymb}
\usepackage{dsfont}
\usepackage{subcaption}
\usepackage{multicol,multirow}
\usepackage[pdftex]{graphicx}   
\usepackage{graphicx}
\usepackage[acronyms, shortcuts]{glossaries}
\usepackage[capitalize,noabbrev,nameinlink]{cleveref}
\crefformat{equation}{(#2#1#3)}
\Crefname{figure}{Fig.}{Fig.}
\Crefname{definition}{Def.}{Def.}
\Crefname{section}{Sec.}{Sec.}
\Crefname{algocfline}{Alg.}{Alg.}
\Crefname{algocf}{line}{lines}
\Crefname{assumption}{Assumption}{Assumptions}
\crefrangeformat{assumption}{Assumptions~#3#1#4-#5#2#6}
\usepackage[ruled,vlined,linesnumbered]{algorithm2e}
\definecolor{princetonorange}{RGB}{232,120,34}

\SetCommentSty{mycommfont}
\SetKwInput{KwInput}{Input}                
\SetKwInput{KwOutput}{Output}              
\usepackage{cases}
\usepackage{xspace}
\usepackage{booktabs}
\usepackage{siunitx}
\usepackage{enumerate}
\usepackage[%
    style=ieee,
    sorting=none,
    natbib=true,  
    backend=biber,
    sortcites=true,
    doi=false,
    url=false,
    giveninits=true,
    maxcitenames=4, 
    minbibnames=3, 
    maxbibnames=4, 
    hyperref
]{biblatex}

\usepackage{balance}

\input{notation}

\IEEEoverridecommandlockouts
\bibliography{references}

\graphicspath{{./figures/}}    
\DeclareGraphicsExtensions{.pdf,.png,.jpg,.jpeg}

\usepackage[textfont=md,font=footnotesize]{caption}

\usepackage{ifthen}
\newboolean{include-notes}
\setboolean{include-notes}{true}

\usepackage{lipsum}

\newbibmacro{string+doiurl}[1]{
    \iffieldundef{doi}{
        \iffieldundef{url}{#1}{\href{\thefield{url}}{#1} }
    }{\href{https://dx.doi.org/\thefield{doi}}{#1}}
}
\DeclareFieldFormat{title}{\usebibmacro{string+doiurl}{\mkbibemph{#1}}}
\DeclareFieldFormat[article,incollection,techreport,inproceedings,book,misc]%
    {title}%
    {\usebibmacro{string+doiurl}{\mkbibquote{#1}}}

\renewbibmacro{in:}{}
\AtEveryBibitem{%
    \clearlist{language}%
    \clearfield{pages}%
    \clearfield{isbn}%
    \clearfield{issn}%
    \clearfield{month}%
    \clearfield{day}
}

\title{\LARGE \bf Back to the Future:\\Efficient, Time-Consistent Solutions in Reach-Avoid Games}
\author{
Dennis R. Anthony, \revise{Duy P. Nguyen,} David Fridovich-Keil, and Jaime F. Fisac
\thanks{
D. R. Anthony is with the Dept. of Mechanical and Aerospace Engineering, Princeton University. D. Fridovich-Keil is with the Dept. of Aerospace Engineering and Engineering Mechanics, UT Austin. \revise{D. P. Nguyen and} J. F. Fisac are with the Dept. of Electrical and Computer Engineering, Princeton University. Correspondence to \href{mailto:dennisra@princeton.edu}{\texttt{dennisra@princeton.edu}}. 
}%
}

\begin{document}

\maketitle
\thispagestyle{empty}
\pagestyle{empty}

\begin{abstract}
\input{abstract}
\end{abstract}

\input{intro}

\input{related}

\input{formulation}

\input{methods}

\input{experiments}

\input{variations}

\input{conclusion}

\section*{Acknowledgments}
The authors sincerely thank Haimin Hu for his assistance with graphical trajectory visualization.

\newpage
\balance
\printbibliography

\end{document}

%% file: notation.tex
\newcommand{\mc}{\mathcal}
\newcommand{\mbf}{\mathbf}
\newcommand{\mbb}{\mathbb}

\newcommand{\reals}{\mbb{R}}
\newcommand{\naturals}{\mbb{N}}
\newcommand{\ball}{\mathcal{B}}
\newcommand{\targetfn}{\ell}
\newcommand{\failurefn}{g}

\newcommand{\targetset}{\mc{T}}
\newcommand{\failureset}{\mc{F}}
\newcommand{\tvar}{t}
\newcommand{\treact}{\tvar_{\textnormal{react}}}
\newcommand{\tinit}{0}
\newcommand{\taux}{\tau}
\newcommand{\tbis}{s}

\newcommand{\pinch}{c}

\newcommand{\marginfn}{m}
\newcommand{\state}{x}

\newcommand{\bstate}{\mbf{x}}
\newcommand{\traj}{\bstate}
\newcommand{\xdim}{n}
\newcommand{\xset}{\mc{X}}
\newcommand{\control}{u}
\newcommand{\bcontrol}{\mbf{\control}}
\newcommand{\controlsignal}{\bcontrol}
\newcommand{\udim}{m}
\newcommand{\uset}{\mc{U}}
\newcommand{\feedback}{K}
\newcommand{\feedforward}{k}
\newcommand{\bfeedback}{\mathbf{\feedback}}
\newcommand{\bfeedforward}{\mathbf{\feedforward}}

\newcommand{\dyn}{f}
\newcommand{\cost}{J}

\newcommand{\horizon}{T}

\newcommand{\strategy}{\gamma}
\newcommand{\strategyset}{\Gamma}
\newcommand{\bstrategy}{\boldsymbol{\strategy}}

\newcommand{\numplayers}{N}
\newcommand{\regularization}{\eta}

\newtheorem{definition}{Definition}
\newtheorem{theorem}{Theorem}

\newcommand{\david}[1]%
    {\textcolor{orange}{\textbf{DFK: #1}}}
    
\newcommand{\dennis}[1]%
    {\textcolor{teal}{\textbf{DA: #1}}}
    
\newcommand{\jaime}[1]%
    {\textcolor{magenta}{\textbf{JFF: #1}}}
    
\newcommand{\revise}[1]%
    {#1}
    
\newcommand{\example}[1]%
{
\textbf{Running example:}
\textit{#1}
}

\newtheorem{remark}{Remark}

\glsdisablehyper
\newacronym[longplural={\revise{P}artially \revise{O}bservable Markov \revise{D}ecision \revise{P}rocesses}]{pomdp}{POMDP}{\revise{P}artially \revise{O}bservable Markov \revise{D}ecision \revise{P}rocess}
\newacronym[longplural={Markov \revise{D}ecision \revise{P}rocesses}]{mdp}{MDP}{Markov \revise{D}ecision \revise{P}rocess}
\newacronym[longplural={Q Markov \revise{D}ecision \revise{P}rocesses}]{qmdp}{QMDP}{Q Markov \revise{D}ecision \revise{P}rocess}
\newacronym{dpo}{DPO}{direct policy optimization}
\newacronym{ukf}{UKF}{unscented Kalman filter}
\newacronym{bsp}{BSP}{belief space planning}
\newacronym{soc}{SOC}{stochastic optimal control}
\newacronym{ours}{DFOC}{distributional feedback optimal control}
\newacronym{admm}{ADMM}{alternating direction method of multipliers}
\newacronym{sqp}{SQP}{\revise{S}equential \revise{Q}uadratic \revise{P}rogramming}
\newacronym{lq}{LQ}{\revise{L}inear-\revise{Q}uadratic}
\newacronym{lqr}{LQR}{\revise{L}inear-\revise{Q}uadratic \revise{R}egulator}
\newacronym{ilq}{ILQ}{\revise{I}terative \revise{L}inear-\revise{Q}uadratic}
\newacronym{ilqr}{ILQR}{\revise{I}terative \revise{L}inear-\revise{Q}uadratic \revise{R}egulator}
\newacronym{mppi}{MPPI}{model predictive path integral}
\newacronym[longplural={cross entropy methods}]{cem}{CEM}{cross entropy method}
\newacronym{mpc}{MPC}{\revise{M}odel \revise{P}redictive \revise{C}ontrol}
\newacronym{smpc}{SMPC}{scenario-based model predictive control}
\newacronym{mbrl}{MBRL}{model-based reinforcement learning}
\newacronym{mcts}{MCTS}{Monte Carlo tree search}
\newacronym{pde}{PDE}{\revise{P}artial \revise{D}ifferential \revise{E}quation}
\newacronym{hj}{HJ}{Hamilton-Jacobi}

%% file: abstract.tex
We study the class of reach-avoid dynamic games in which multiple agents interact noncooperatively, and each wishes to satisfy a distinct target criterion while avoiding a failure criterion.
Reach-avoid games are commonly used to express safety-critical optimal control problems found in mobile robot motion planning.
Here, we focus on finding \emph{time-consistent} solutions, in which future motion plans remain optimal even when a robot diverges from the plan early on due to, e.g., intrinsic dynamic uncertainty or extrinsic environment disturbances.
Our main contribution is a computationally-efficient algorithm for multi-agent reach-avoid games which renders time-consistent solutions for all players.
We demonstrate our approach in \revise{two- and three-player} simulated driving scenario\revise{s}, \revise{in which our method provides safe control strategies for all agents.}

%% file: intro.tex
\section{Introduction}
\label{sec:intro}

\begin{figure}
    \centering
    \includegraphics[width=\linewidth]{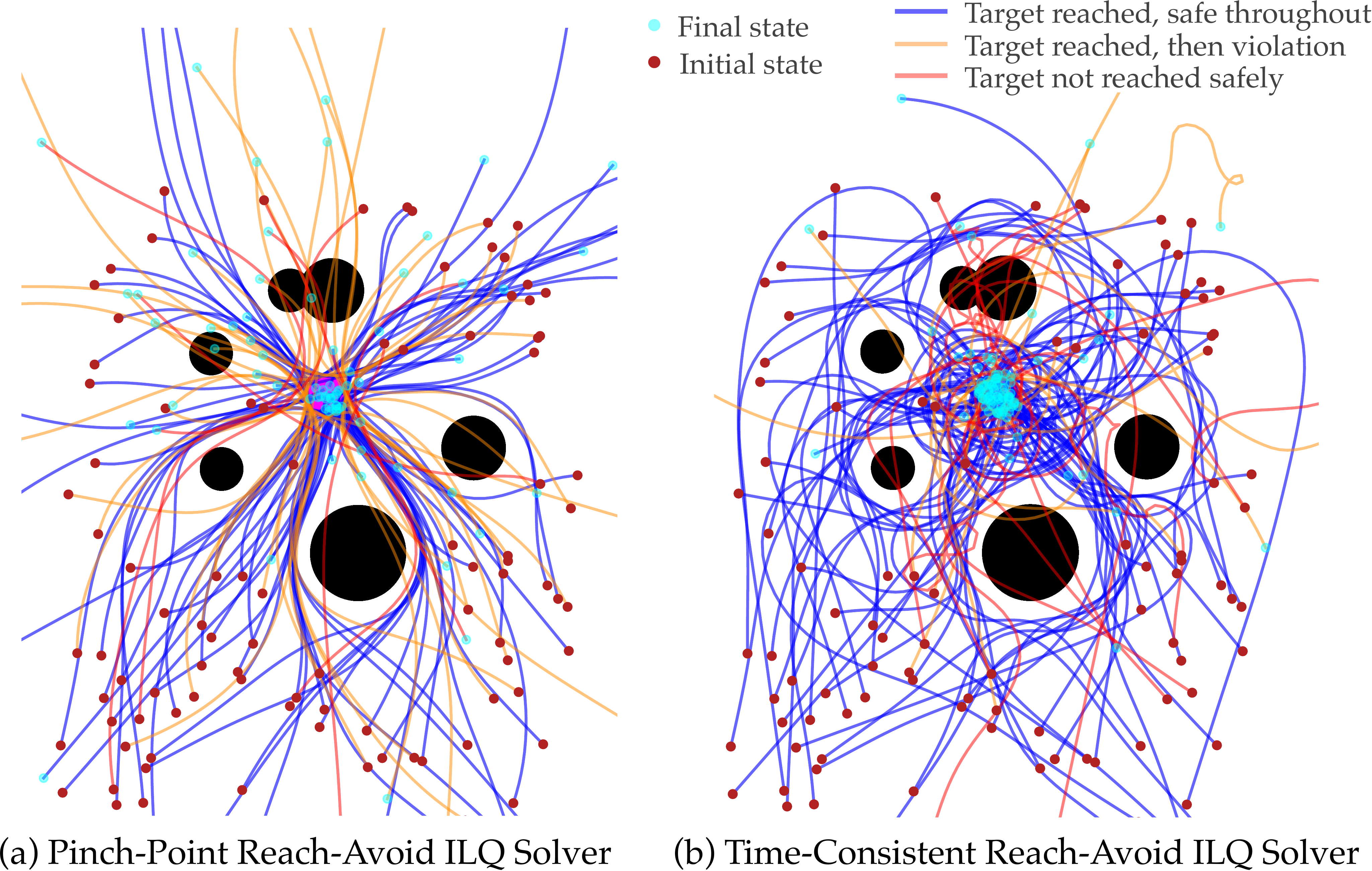}
    \caption{
    Comparison of \ac{ilq} reach-avoid methods evaluated at 100 random initial states. Here, a planar vehicle is \revise{required} to reach the the center position while avoiding the black obstacles.
    \revise{The existing pinch-point method ignores events that occur after the vehicle first reaches the target, and often leads the vehicle into the failure set.
    Conversely, our proposed time-consistent method  
    yields trajectories that continue to seek the target while attempting to remain safe at all times.}
    \vspace{-5mm}
    }
    \label{fig:one-player-comparison}
\end{figure}

Consider a canonical motion planning problem, in which a mobile robot must reach a pre-specified goal condition while avoiding obstacles or other hazards.
There are a wide variety of approaches for solving these types of problems, including randomized graph planning, smooth (non)convex optimization, and dynamic programming.
In many ways, such methods are interchangeable in that they produce an optimal robot trajectory which achieves the desired outcome without violating any safety constraints.
This paper concentrates on a subtle distinction, however, which is intimately related to solution robustness. 
That is, we focus on the \emph{time consistency} of solutions to these reach-avoid problems, as illustrated in~\cref{fig:one-player-comparison}.

Informally, a time-consistent solution is one which is still optimal despite variation early on in the robot's trajectory.
Time-\emph{in}consistent plans are thus highly susceptible to uncertainty in robot motion and environment structure, such as the location of obstacles.
This susceptibility is well-known in the motion planning community, and necessitates rapid re-planning to account for the latest sensor observations.

Established methods for identifying time-consistent solutions to reach-avoid problems amount to solving a \ac{pde} whose dimensionality corresponds to the dimension of the robot's state space.
Unfortunately, this requires exponential computation time and memory in general, relegating reach-avoid verification to offline operation.
Recent progress establishes an efficient technique for computing real-time local solutions in the avoid-only case \cite{fridovich2021approximate}.
Unfortunately, however, this approach is time-inconsistent.
In that context, the contributions of this paper are threefold: \begin{enumerate}
    \item \emph{Efficient reach-avoid algorithm.} We propose a novel, computationally efficient technique for finding approximate solutions to reach-avoid problems.
    \item \emph{Time consistency.} We modify this approach such that it finds (locally) time-consistent control strategies, in which later control inputs are still optimal despite suboptimal behavior earlier in the trajectory.
    \item \emph{Multi-agent, game-theoretic settings.} Many modern robotic systems operate in the presence of other agents with potentially competing objectives (e.g., traffic). We derive and demonstrate our methods in the setting of multi-player reach-avoid games.
\end{enumerate}


%% file: related.tex
\section{Prior Work}
\label{sec:related}



Robot motion planning methods aim to guide a robot from an initial configuration to a desired one while avoiding undesired conditions such as collisions with obstacles.
These two temporal logic properties are commonly referred to, respectively, as \emph{liveness} (a desired condition will eventually take place) and \emph{safety} (undesired conditions will never take place)~\cite{alpern1985defining}.
Historically, such problems have been primarily addressed through two complementary lenses: search and optimization.
Search-based methods \cite{kavraki1996probabilistic, lavalle1998rapidly, karaman2011sampling} construct (potentially randomized) graphs of feasible robot motion, and identify optimal paths therein while enforcing safety
during graph construction by rejecting any candidate nodes or edges that violate the state constraints. 
However, the performance of these methods suffers in high-dimensional problems, making them challenging to use in real-time applications and especially in safety-critical settings.

Continuous optimization-based methods exploit the differentiability of robot dynamics and planning objectives to solve certain motion planning problems more efficiently.
However, since these methods rely upon derivative information, they find local solutions as opposed to the global solutions identified by graph search methods.
Still, optimization-based methods~\cite{Nocedal:2006, jacobson1970differential} are widely used in real-time motion planning and \ac{mpc} \cite{borrelli2017predictive}.

Our method falls within the optimization-based category, but
\revise{uses a level set method \cite{isaacs1954differential} to enforce richer trajectory-wide properties in the form of temporal logic specifications, rather than state and control constraints at individual time instants.}
With level set methods, reach-avoid objectives can be constructed so that the sign of a scalar outcome determines whether the safety and liveness conditions have been met~\cite{bokanowski2010reachability,fisac2015reachavoid}.
This continuous scalar encoding induces a dynamic programming relation in the form of a \ac{hj} \ac{pde} or variational inequality, whose viscosity solution \cite{evans1984differential} determines the outcome of the game when all players choose their control inputs optimally.
This \acrlong{hj} equation is usually solved numerically \cite{mitchell2008flexible}; however computation scales poorly with state dimension due to the ``curse of dimensionality''~\cite{bellman1956dynamic}.

Recently, however, differential dynamic programming methods \cite{jacobson1970differential}, such as the \acrfull{ilqr} \cite{van2014iterated} have proven extremely successful in trajectory optimization problems, and have been extended to multi-agent settings~\cite{fridovich2020efficient, di2019newton}.
These ideas have motivated the investigation of a efficient level set formulation for the safety (avoid-only) case \cite{fridovich2021approximate}.
As we prove in \cref{sec:methods}, this approach unfortunately results in a time-inconsistent control strategy.
In this work, we develop a novel approach which retains the computational advantages of prior work while yielding a more robust, time-consistent solution.

%% file: formulation.tex
\section{Problem Formulation}
\label{sec:formulation}


\subsection{Reach-Avoid Problem}
We begin by considering a general reach-avoid optimal control problem, in which we seek to drive the state ${\state_\tvar \in \xset\subseteq\reals^\xdim}$ of a dynamical system into a (possibly dynamic) target region $\targetset_\tvar\subset\xset$ at some future time $\tvar$ while avoiding a (possibly time-varying) failure region $\failureset_\tvar \subset \xset$ at all previous times.
Let the discrete-time dynamics of the system be given by
$\state_{\tvar+1} = \dyn_\tvar(\state_\tvar,\control_\tvar)$,
with $\control_\tvar\in\uset\in\reals^\udim$ denoting the control input at time $\tvar$.

The reach-avoid condition for a generic state trajectory ${\traj := (\state_\tinit,\dots,\state_\horizon)}$ can be expressed as:
\begin{equation}
    \label{eq:reach-avoid-condition}
     \exists\tvar\in\{\tinit,\dots,\horizon\}: \state_\tvar\in\targetset_\tvar \, \wedge \,
     \forall\taux\in\{\tinit,\dots,\tvar\}: \state_\taux\not\in\failureset_\taux
\end{equation}
We only assume that $\targetset$ is closed and $\failureset$ is open, following the convention in the literature~\cite{bokanowski2010reachability,fisac2015reachavoid}; no other assumptions (convexity, connectedness, etc.) are required.
We can then characterize the target and failure sets using two (possibly time-varying) Lipschitz continuous \emph{margin functions}, ${\targetfn_\tvar, \failurefn_\tvar:\xset\to\reals}$, as follows:
\begin{equation}
    \label{eq:level-set}
    \targetfn_\tvar(\state_\tvar)\le 0 \iff \state_\tvar\in\targetset_\tvar\,,\;
    \failurefn_\tvar(\state_\tvar) > 0 \iff \state_\tvar\in\failureset_\tvar\,.
\end{equation}
This characterization is the basis of the level set method, which allows us to encode the reach-avoid condition~\cref{eq:reach-avoid-condition} through the following scalar objective, which we seek to minimize by our choice of control strategy:
\begin{equation}
    \label{eq:reach-avoid-objective}
     \cost_\tbis(\traj) = \min_{\tvar \in \{\tbis,...,\horizon\}} \max \{\targetfn_\tvar(\state_\tvar), \max_{\taux \in \{\tbis,...,\tvar\}} \failurefn_\taux(\state_\taux) \}\,.
\end{equation}
Note that we have expressed this objective-to-go along the time horizon $\{\tbis,\dots,\horizon\}$ for an arbitrary $\tbis\in\{\tinit,\dots,\horizon\}$: states visited by trajectory $\traj$ prior to $\tbis$ are neglected by $\cost_\tbis$.
In general, we seek solutions for $\tbis = \tinit$, but as we shall see, subproblems for other $\tbis > \tinit$ will allow us to define and derive more robust \emph{time-consistent} solutions.

\subsection{Time Consistency}

Time consistency expresses the desirable property that an optimal control strategy \emph{remains optimal} for sub-problems beginning at intermediate times along its execution despite earlier suboptimal actions. In other words, the robot has no incentive to deviate from a time-consistent strategy.

The traditional definitions of time consistency in the optimal control and dynamic game theory literature (cf. \cite[Ch. 5]{basar1998dynamic}) are specific to problems with time-additive costs.
Here, we extend \revise{these definitions} to reach-avoid problems \revise{in which margin function values at early times can dominate overall cost}.
\revise{The extensions below preserve the spirit of the time-additive definitions}, and continue to encode the rationality of later control decisions regardless of state history.

The notion of time consistency can be applied to a variety of control strategy classes, such as state feedback policies and open-loop control signals.
Let $\strategy\in\strategyset$ be the allowed control strategies along the time horizon $\{\tinit,\dots,\horizon\}$, and 
denote by $\strategy_\tbis\in\strategyset_\tbis$ their truncation
to the time horizon $\{\tbis,\dots,\horizon\}$ for $\tbis\in\{\tinit,\dots,\horizon\}$.

\begin{definition}[Weak time consistency]
\label{def:weak-time-consistency}
An optimal control strategy $\strategy^*\in\strategyset$ for the reach-avoid objective $\cost_\tinit$
is weakly time-consistent if its truncation to $\{\tbis,\dots,\horizon\}$
remains optimal for the truncated reach-avoid problem with objective $\cost_\tbis$
that results after
applying $\strategy^*$ on $\{\tinit,\dots,\tbis-1\}$.
That is, given the optimal strategy $\strategy^*$ (which, by assumption, achieves the minimum $\cost_\tinit$ from all initial conditions $\state_\tinit\in\xset_\tinit$), consider any alternative strategy $\strategy\in\strategyset$ that is identical to $\strategy^*$ at times $\tvar < \tbis$ and may differ thereafter. Then the following inequality must hold for every such $\strategy$:
\begin{equation*}
    \cost_\tbis\big(\traj^{\strategy^*}_{\tinit, \state_\tinit}\big) \le \cost_\tbis\big(\traj^{\strategy}_{\tinit, \state_\tinit}\big), \forall \tbis \in \{\tinit, \dots, \horizon\},\forall \state_\tinit\in\xset_\tinit\,\revise{,}
\end{equation*}
\revise{w}ith $\traj^\strategy_{\tinit, \state_\tinit}$ denoting the state trajectory resulting from applying strategy $\strategy$ from initial conditions $\state_\tinit$ at time $\tinit$.
A solution $\strategy^*$ that is not weakly time-consistent is \emph{time-inconsistent}.
\end{definition}

\begin{definition}[Strong time consistency]
\label{def:strong-time-consistency}
An optimal control strategy $\strategy^*\in\strategyset$, defined for times $\tvar \in \{\tinit, \dots, \horizon\}$ is strongly time-consistent if it remains optimal for the truncated reach-avoid problem when restricted to begin at an intermediate time $\tbis$ \emph{from an arbitrary state} $\tilde \state$.
That is, consider any other strategy $\strategy$ defined at subsequent times $\tvar \in \{\tbis, \dots, \horizon\}$. Then the following inequality must hold for every such $\strategy$:
\begin{equation*}
    \cost_\tbis\big(\traj_{\tbis,\tilde\state}^{\strategy^*}\big) \le \cost_\tbis\big(\traj_{\tbis,\tilde\state}^{\strategy}\big), \forall \tbis \in \{\tinit, \dots, \horizon\},\, \forall\tilde\state\in\xset.
\end{equation*}
A strongly time-consistent solution $\strategy^*$ is also weakly time-consistent; the converse is not in general true.
\end{definition}

\begin{remark}
    In some cases, we may only be interested in the reach-avoid problem from a known initial state, in which case $\xset_\tinit :=\{\state_\tinit\}$. This is commonly the case for online trajectory optimization schemes, like the one proposed here.
\end{remark}

\subsection{General-Sum Multi-Player Reach-Avoid Games}

For the remainder of this paper, we consider the more general setting of an $\numplayers$-player game in which each player's objective, denoted $\cost^i$ (with target and failure margin functions $\targetfn^i$ and $\failurefn^i$), is of the form \cref{eq:reach-avoid-objective}, and the overall state of the system $\state$ evolves jointly with all players' inputs $\control^i, i \in \{1,\dots,\numplayers\}$, i.e., $\state_{\tvar+1} = \dyn_\tvar(\state_\tvar, \control^1_\tvar, \dots, \control^\numplayers_\tvar)$.
In this multi-player setting, time-consistency is defined analogously to \cref{def:weak-time-consistency,def:strong-time-consistency}, where ``solution'' is understood to refer to the desired game-theoretic equilibrium concept%
\revise{, as in \cite[Ch. 5, footnote 44]{basar1998dynamic}}---in our case, a local Nash equilibrium. 

\begin{definition} \emph{Local Nash Equilibrium \cite[Def. 1]{ratliff2016characterization}}
\label{def:lne}
    A tuple of player strategies $\bstrategy^*:=(\strategy^{i*})_{i=1}^\numplayers$ is a local Nash equilibrium if there exists a neighborhood of strategies $\tilde\strategyset^i\subset\strategyset^i$ for each player (with $\strategy^{i*}\in\tilde\strategyset^i$) within which that player has no incentive to unilaterally deviate from strategy $\strategy^{i*}$.
    That is, the following must hold for every player $i$:
    \begin{multline*}
        \exists \tilde\strategyset^i\subset\strategyset^i: \strategy^{i*}\in\tilde\strategyset^i \quad \wedge \\
        \cost^i_\tinit\big(\traj_{\state_\tinit}^{\bstrategy^*}\big) \le \cost^i_\tinit\big(\traj_{\state_\tinit}^{(\bstrategy^{\neg i*},\strategy^i)}\big),\forall \strategy^i\in\tilde\strategyset^i,\forall \state_\tinit\in\xset_\tinit,
    \end{multline*}
    where $(\bstrategy^{\neg i*},\strategy^i)$ represents a tuple of strategies where all players except for $i$ follow $\bstrategy^*$ and player $i$ uses $\strategy^i$.
\end{definition}

%% file: methods.tex
\section{Technical Approach}
\label{sec:methods}


Recent work \cite{fridovich2021approximate} develops an efficient method for solving reachability (reach-only or avoid-only) optimal control problems to a locally optimal feedback policy, as well as multi-player reachability games to a local feedback Nash equilibrium.
This approach is based upon earlier work for computing local Nash equilibria in general-sum dynamic games with time-additive costs using \ac{ilq} approximations \cite{fridovich2020efficient}, theoretical convergence analysis of which is given in \cite{laine2021computation}.
In the reachability setting of \cite{fridovich2021approximate}, each player's objective is of the form
     ${\cost_\tbis(\traj) = \min_{\tvar \in \{\tbis,...,\horizon\}} \targetfn_\tvar(\state_\tvar)}$,
which can be seen as a special case of \cref{eq:reach-avoid-objective} for $\failurefn_t(\cdot)\equiv-\infty$.

Each \ac{lq} approximation of the game is formed by first finding the time $\taux$ at which the above reachability objective attains its minimum for the current iteration's state trajectory, and then minimizing (liveness) the value of $\targetfn_{\taux}(x_{\taux})$ (alternatively maximizing for safety problems).
The resulting \ac{lq} game is in the standard time-additive cost form, with only one nonzero state cost term along the entire horizon (namely, at $\tvar=\taux$),
since the remaining time steps do not locally contribute to $\cost_\tbis(\traj)$ within a neighborhood of the current trajectory iterate.  
To reflect this structure, we refer to this approach as the ``pinch-point method.''

\subsection{Pinch-Point Reach-Avoid}
\label{sec:pinch-point-reach-avoid}

Our first contribution is to extend the pinch-point method to reach-avoid problems with player objectives in the form of \cref{eq:reach-avoid-objective}. 
The extension is conceptually straightforward, but the more complex form of the reach-avoid objective  \cref{eq:reach-avoid-objective} requires carefully keeping track of both $\targetfn$ and $\failurefn$ to resolve which margin function, and at what time, is determining the value of $\cost_\tinit(\traj)$ along the current trajectory iterate $\traj$.

The proposed extension is summarized in \cref{alg:pinch-point-subroutine}, which forms a subroutine of the iterative \ac{lq} game procedure of \cref{alg:main-alg} from \cite{fridovich2020efficient} and \cite{fridovich2021approximate}. 

\input{main_alg}
\input{pinch_point_subroutine}

In \cref{alg:main-alg}, we iteratively refine the set of strategies for all players, $\bstrategy = (\strategy^1, \dots, \strategy^\numplayers$) by solving \ac{lq} games which locally approximate the underlying reach-avoid game where each player's objective is of the form \cref{eq:reach-avoid-objective}.
The reach-avoid pinch-point method in \cref{alg:pinch-point-subroutine} (extending that of \cite{fridovich2021approximate}), forms a standard, time-additive \ac{lq} approximation of the type considered in standard references such as \cite[Ch. 6]{basar1998dynamic}.
In particular, we find a linear approximation to the dynamics about the current trajectory iterate $\traj$ (iteration index suppressed for clarity) and, critically, a time-additive quadratic approximate objective of the form
\begin{equation}
    \label{eq:time-additive-quadratic-approx}
    \tilde \cost^i_\tinit = \frac{1}{2} \sum_{\tvar = \tinit}^\horizon \Big(\big(\state_\tvar^\top Q^i_\tvar + 2 q^{i \top}_\tvar\big) \state_\tvar + \big(\control^{i\top}_\tvar R^i_\tvar + 2 r^{i \top}_\tvar\big) \control^i_\tvar\Big)\,.
\end{equation}
Here, $Q^i_\tvar$ and $q^i_\tvar$ are a quadratic approximation of the original reach-avoid objective evaluated at the pinch-point or critical time, i.e. the unique time $\taux$ at which the trajectory's cost $\cost^i_\tinit$ evaluates to either the target margin function $\targetfn_\taux(\state_\taux)$ or the failure margin function $\failurefn_\taux(\state_\taux)$.
Following \cite{fridovich2021approximate}, we also introduce a small quadratic regularization into the original reach-avoid objective of each player, i.e. $\cost^i_\tinit \gets {\cost^i_\tinit + \regularization \sum_\tvar \|\control^i_\tvar\|^2}, \regularization>0$ to make the problem numerically well-posed.
$R^i_\tvar$ and $r^i_\tvar$ constitute a quadratic approximation to this regularization term about input trajectory $\controlsignal$.
Thus equipped, \cref{alg:pinch-point-subroutine} leverages established methods \cite[Cor. 6.1]{basar1998dynamic} to compute the unique feedback Nash equilibrium of the \ac{lq} game with time-additive objectives \cref{eq:time-additive-quadratic-approx}.

\looseness=-1
Unfortunately, the pinch-point solutions computed using \cref{alg:pinch-point-subroutine}---and by extension the safety-only approach of \cite{fridovich2021approximate}---are time-inconsistent.

\begin{theorem}[Pinch-point solutions are time-inconsistent]
\label{thm:pinch}
Suppose that the pinch-point method of \cite{fridovich2021approximate} converges to a game trajectory $\traj$ with strategies $\bstrategy^*$.
These strategies are generally time-inconsistent and need not satisfy either \cref{def:weak-time-consistency} or \cref{def:strong-time-consistency}.\\
\begin{proof}
We offer a counterexample, which is illustrated in \cref{fig:one-player-comparison} and described in further detail below. 
\end{proof}
\end{theorem}


This result is illustrated for a single-player setting in \cref{fig:one-player-comparison}.
Here, we see that a vehicle---modeled with the dynamics given in \cref{eq:bicycle-dynamics}---seeks to \emph{reach} the small magenta disk while \emph{avoiding} the larger black disk.
The target margin function $\targetfn$ and failure margin function $\failurefn$ encode signed distance to the boundaries of these sets, respectively.
The pinch-point strategy avoids the failure set and reaches the target set, but thereafter applies no input and drifts.

This is because \emph{any} trajectory which avoids the failure set and eventually reaches the target set has \emph{the same} cost $\cost_\tinit$.
Consider a time $\taux$ for which the vehicle has already passed the target set, and the associated cost-to-go $\cost_\taux$.
For a sufficiently long time horizon $\horizon$, the cost-to-go associated with the remainder of the trajectory $\state_\tvar, \taux \le \tvar \le \horizon$, could be reduced by turning the vehicle back toward the target set so that it enters once again.
Thus, $\cost_\taux$ is not minimized by the pinch-point solution and, by \cref{def:weak-time-consistency} it is time-inconsistent.

\subsection{Time-Consistent Reach-Avoid}
\label{sec:time-consistent-reach-avoid}

Our second, and more substantial, contribution is a \emph{locally strongly time-consistent} algorithm for reach-avoid problems.
This approach shares the \ac{ilq} structure of \cref{alg:main-alg}, but we derive a new time-consistent solution for the \ac{lq} case, given in \cref{alg:time-consistent-subroutine}.
As before, we begin by identifying times at which the reach-avoid objective $\cost^i$ takes the value of either the target margin function $\targetfn^i$ or failure margin function $\failurefn^i$. 
However, where \cref{alg:pinch-point-subroutine} records only a single so-called pinch point which determines the value of $\cost^i_\tinit$, here we record a set of critical times $\{\taux_j\}$ and associated margin functions $\marginfn^i_{\taux_j, j}(\cdot)$---which are either $\targetfn^i_{\taux_j, j}(\cdot)$ or $\failurefn^i_{\taux_j, j}(\cdot)$---such that
\begin{equation}\label{eq:obj_crit}
    \cost^i_\tvar(\cdot) = \marginfn^i_{\taux_j, j
    }(\cdot), \forall \taux_{j-1} < \tvar \le \taux_j\,,
\end{equation}
presuming that the critical times $\taux_j$ are indexed by $j \in \naturals$.

After computing the set of critical times for each player $i$, \cref{alg:time-consistent-subroutine} solves the reach-avoid \ac{lq} game in which each player's cost-to-go at time $\tvar$ is a quadratic approximation of the associated margin function $\marginfn^i_{\taux_j, j}(\cdot)$ for the subsequent critical time $\taux_j$.
As in \cref{eq:time-additive-quadratic-approx}, the quadratic approximation $\tilde \cost^i_\tvar$ has the form
\begin{multline}
    \label{eq:time-consistent-lq-objective}
    \tilde \cost^i_\tvar = \frac{1}{2} \Bigg(\big(\state_{\taux_j}^\top Q^i_{\taux_j} + 2 q^{i \top}_{\taux_j}\big) \state_{\taux_j} + \sum_{\tbis = \tvar}^{\taux_j} \Big(\big(\control^{i \top}_\tbis R^i_\tbis + r^{i \top}_\tbis\big) \control^i_\tbis\Big)\Bigg),\\
    \forall \taux_{j-1} < \tvar \le \taux_j\,,
\end{multline}
Here, $Q^i_{\taux_j}$ and $q^i_{\taux_j}$ constitute a quadratic approximation of time $\tvar$'s active margin function $\marginfn^i_{\taux_j}$ about the current trajectory iterate $\traj$, and we incorporate the same control regularization term as in \cref{eq:time-additive-quadratic-approx} in the terms $R^i_\tbis$ and $r^i_\tbis$.

\input{time_consistent_subroutine}

We find a time-consistent solution of this \ac{lq} game by modifying the standard backward recursion used to solve time-additive \ac{lq} games \cite[Cor. 6.1]{basar1998dynamic}.
That is, we represent the cost-to-go from a given state with the matrix-vector pair $(Z^i_\tvar, z^i_\tvar)$, where $\tilde \cost^i_\tvar(\state_\tvar) = \frac{1}{2}\big(\state_\tvar^\top Z^i_\tvar + z^{i\top}_\tvar\big)\state_\tvar$.
Whenever $\tvar$ is a critical time $\taux_j$, we fix $(Z^i_\tvar, z^i_\tvar) = (Q^i_{\taux_j}, q^i_{\taux_j})$, and at intervening times we follow the standard time-additive coupled Riccati recursion \cite{basar1998dynamic} setting $Q^i_\tvar$ and $q^i_\tvar$ to zero since no margin functions are active during this interval.

This procedure explicitly considers the sequence of \ac{lq} games whose objectives encode the subsequent critical time and associated margin function.
Since feedback Nash solutions to \ac{lq} games are strongly time-consistent \cite[Chapter 6]{basar1998dynamic}, our approach identifies time consistent solutions as well.
To make this precise, we define local time consistency for a single agent, in line with \cref{def:weak-time-consistency} and \cref{def:strong-time-consistency}.

\begin{definition}[Local strong time consistency]
\label{def:local-strong-time-consistency}
A locally optimal control strategy $\strategy\in\strategyset$, defined for times ${\tvar \in \{\tinit, \dots, \horizon\}}$ is \emph{locally} strongly time-consistent if it remains \emph{near}-optimal for the truncated reach-avoid problem restricted to begin at an intermediate time $\tbis$ from an arbitrary state $\tilde \state$ within a neighborhood of the optimal trajectory at that time.
\looseness=-1
That is, for any $\epsilon > 0$, there exist $\delta_\state, \delta_\strategy>0$ such that, for each time step ${\tbis \in \{\tinit, \dots, \horizon\}}$,
\begin{equation*}
    \cost_\tbis\big(\traj_{\tbis,\tilde\state}^{\strategy^*}\big) \le \cost_\tbis\big(\traj_{\tbis,\tilde\state}^{\strategy}\big) + \epsilon,\;
    \forall \tilde\state \in\ball(\state^*_\tbis,\delta_\state),
    \forall \strategy\in\ball(\strategy^*,\delta_\strategy).
\end{equation*}
\end{definition}

\vspace{0.5em}
As before, the definition extends to multi-player decision problems by interpreting ``optimality'' as per the desired game solution concept (e.g. feedback Nash equilibrium).
We now state our main theorem regarding the time-consistency of \cref{alg:main-alg} and \cref{alg:time-consistent-subroutine}.

\begin{theorem}
Let the solution $\bstrategy^*$ returned by \cref{alg:main-alg} using the subroutine provided in \cref{alg:time-consistent-subroutine} be 
locally optimal under the feedback Nash equilibrium concept.
Then, $\bstrategy^*$
exhibits (feedback Nash) local strong time consistency.\\
\begin{proof}
    By construction, the affine feedback policies produced by \cref{alg:time-consistent-subroutine} have constitute a feedback Nash equilibrium to the \ac{lq} game defined by the derivatives of players' reach-avoid objectives $\cost_0^i$.
    Further, the stage-wise feedback strategies $(\feedback^i_\tvar, \feedforward^i_\tvar)_{i=1}^\numplayers$ at each time $\tvar$ are optimal for the objective $\cost_t^i$, as given by \cref{eq:obj_crit}.
    Therefore, these strategies are \emph{globally} strongly time-consistent for the \ac{lq} game that locally approximates the overall dynamic game.
    By hypothesis, $\bstrategy^*$ is a local feedback Nash equilibrium, indicating that there exists $\delta_\strategy>0$ such that, at each time ${\tvar\in\{\tinit,\dots,\horizon\}}$, 
    $\cost^i_\tvar\big(\traj_{\state_\tvar^*}^{\bstrategy^*}\big) \le \cost^i_\tvar\big(\traj_{\state_\tvar^*}^{(\bstrategy^{\neg i*},\strategy^i)}\big), \forall \strategy^i\in\ball(\strategy^{i*},\delta_\strategy)$.
    Therefore, due to the smoothness of players' costs and the continuous dependence of trajectories upon initial conditions (cf.~\cite[Thm. 3.23]{sastry1991nonlinear}), for any $\epsilon>0$ there exists a $\delta_\state>0$ such that, for all ${\tilde\state \in\ball(\state^*_\tvar,\delta_\state)}$, the inequality of \cref{def:local-strong-time-consistency} holds.
\end{proof}
\end{theorem}

We showcase this result in \cref{fig:one-player-comparison}, which
\revise{compares the trajectories computed by the \ac{ilq} reach-avoid scheme (\cref{alg:main-alg}) using the pinch-point (\cref{alg:pinch-point-subroutine}) and time-consistent (\cref{alg:time-consistent-subroutine}) subroutines}.
In the pinch-point formulation of \cref{sec:pinch-point-reach-avoid}, all trajectories that passed through the target while avoiding the obstacle was indistinguishably optimal.
Here, however, any trajectory which terminates at the target while always avoiding the obstacle is indistinguishable.
This set of trajectories is precisely that which satisfies the reach-avoid condition \cref{eq:reach-avoid-condition} on subsets of the time horizon $\tvar \in \{\tbis, \dots, \horizon\}$ and is hence time-consistent.


\subsection{Computational Efficiency \revise{and Optimality of Solutions}}
\label{sec:efficiency}

\revise{While the proposed time-consistent approach introduces an additional Boolean check in Line 5 of \cref{alg:time-consistent-subroutine},}
both time-(in)consistent subroutines have asymptotic complexity commensurate with that of the standard coupled Riccati solution to time-additive \ac{lq} games, which scales with the cube of the number of players and the state dimension, and is linear in the time horizon~\cite{fridovich2020efficient,basar1998dynamic}.
This polynomial scaling is equivalent to that of standard second-order solution methods such as Newton's method or \acs{ilqr} in the single-player setting, 
making the family of \ac{ilq} game solution methods
amenable to real-time implementation.
Indeed, experimental evidence in \cref{tbl:stats} suggests that our time-consistent method further improves the iteration complexity of \cref{alg:main-alg}, while also improving safety performance.

\begin{table}
\centering
\begin{tabular}{ c||c|c|c } 
 & target reached (\#) & mean (max) \# iters & safe after target (\#)\\
 \hline
 PP & \textbf{93} & 33.94 (140) & 63 \\ 
 TC & 88 & \textbf{18.6 (77)} & \textbf{78}\\ 
 \hline
\end{tabular}
\caption{Summary statistics of pinch-point (PP) and time-consistent (TC) reach-avoid solvers from 100 trajectories shown in \cref{fig:one-player-comparison} starting from the same 100 randomly sampled initial states. Both methods were terminated as soon as the condition $\cost_0\le 0$ was met (target reached without \emph{previous} constraint violations; if allowed to iterate further, TC finds 84 (instead of 78) trajectories that remain safe for all time.}
\label{tbl:stats}
\vspace{-.5cm}
\end{table}


%% file: main_alg.tex
\begin{algorithm}[tbp]
\label{alg:main-alg}
\DontPrintSemicolon
\caption{\ac{ilq} reach-avoid algorithm}
    \KwInput{${\state_\tinit}, \horizon, (\targetfn^i_\tvar (\cdot), \failurefn^i_\tvar(\cdot))_{i=1}^\numplayers$}
    Initialize $\strategy^i \gets(\controlsignal^i, \bfeedback^i)$ \tcp*{e.g. open-loop, $\bfeedback^i \equiv 0$}
	\While{not converged}{
        $\traj \gets \text{RolloutTrajectory}\big(\state_\tinit,(\controlsignal^i)_{i=1}^\numplayers\big)$\\
	    $(\tilde\bfeedback^i,\tilde\bfeedforward^i)_{i=1}^\numplayers \gets \textbf{ReachAvoidLQSolve}\big(\traj,(\controlsignal^i)_{i=1}^\numplayers\big)$ \tcp*{either \cref{alg:pinch-point-subroutine} or \cref{alg:time-consistent-subroutine}}
	    $\bstrategy\gets \text{\ac{ilq}StrategyUpdate}\big(\state_\tinit,(\controlsignal^i,\tilde\bfeedback^i,\tilde\bfeedforward^i)_{i=1}^\numplayers\big)$ \tcp*{e.g., the line search method of \cite{laine2021computation}}
    }
	\Return converged strategies $\bstrategy$
\end{algorithm}

%% file: pinch_point_subroutine.tex
\begin{algorithm}[tbp]
\label{alg:pinch-point-subroutine}
\DontPrintSemicolon
\caption{Pinch-point reach-avoid \ac{lq} solver}
    \KwInput{$\traj,\big(\controlsignal^i, \targetfn^i_\tvar (\cdot), \failurefn^i_\tvar(\cdot)\big)_{i=1}^\numplayers$}
    \For(\tcp*[f]{find pinch point $\forall$ players}){$i=1,\dots,\numplayers$}{
        $\cost_{T+1}^i \gets \infty$\;
        \For(\tcp*[f]{dynamic programming}){$\tvar=\horizon,\dots,\tinit$}{
            $\cost_\tvar^i \gets \max\big\{\failurefn_\tvar^i(\state_\tvar), \min \{\cost_{\tvar+1}^i, \targetfn_\tvar^i(\state_\tvar)\}\big\}$\;
            $\pinch^i \gets \begin{cases}
            \big(\tvar,\failurefn_\tvar^i(\cdot)\big) & \textbf{if }\cost_\tvar^i = \failurefn_\tvar^i(\state_\tvar)\\
            \big(\tvar, \targetfn_\tvar^i(\cdot)\big)& \textbf{if }\cost_\tvar^i = \targetfn^i_\tvar(\state_\tvar)\\
            \revise{c^i} & \revise{\textbf{otherwise}}
            \end{cases}$\;
        }
    }
    \tcp{set up LQ game}
    $Q_{\tvar}^i, q_{\tvar}^i \gets \text{Quadratize}\big(\marginfn^i(\cdot)\big)$ \textbf{for} $(\tvar, \marginfn^i) \in \pinch^i$,~\textbf{else}~$0$\;
    $R_\tvar^i,r^i_\tvar \gets\text{Quadratize}\big(\regularization \|\control_\tvar^i\|^2\big),\forall i,\tvar$\;
    $(\tilde\bfeedback^i,\tilde\bfeedforward^i)_{i=1}^\numplayers \gets$~SolveLQGame$\big((\mathbf{Q}^i, \mathbf{q}^i, \mathbf{R}^i, \mathbf{r}^i)_{i=1}^\numplayers\big)$\\
	\Return $(\tilde\bfeedback^i,\tilde\bfeedforward^i)_{i=1}^\numplayers$
\end{algorithm}

%% file: time_consistent_subroutine.tex
\begin{algorithm}[tbp]
\label{alg:time-consistent-subroutine}
\DontPrintSemicolon
\caption{Time-consistent reach-avoid \ac{lq} solver}
    \KwInput{$\traj,\big(\controlsignal^i, \targetfn^i_\tvar (\cdot), \failurefn^i_\tvar(\cdot)\big)_{i=1}^\numplayers$}
    \For(\tcp*[f]{find all critical times}){$i=1,\dots,\numplayers$}{
        $\cost_{T+1}^i \gets \infty$, $\pinch^i\gets\varnothing$\;
        \For(\tcp*[f]{dynamic programming}){$\tvar=\horizon,\dots,\tinit$}{
            $\cost_\tvar^i \gets \max\big\{\failurefn_\tvar^i(\state_\tvar), \min \{\cost_{\tvar+1}^i, \targetfn_\tvar^i(\state_\tvar)\}\big\}$\;
            $\pinch^i \gets \begin{cases}
            \big\{\pinch^i, \big(\tvar,\failurefn_\tvar^i(\cdot)\big)\big\}& \textbf{if }\cost_\tvar^i = \failurefn_\tvar^i(\state_\tvar)\\
            \big\{\pinch^i, \big(\tvar, \targetfn_\tvar^i(\cdot)\big)\big\}& \textbf{if }\cost_\tvar^i = \targetfn^i_\tvar(\state_\tvar)\\
            \revise{c^i} & \revise{\textbf{otherwise}}
            \end{cases}$
        }
    }
    \tcp{set up LQ game}
    $Q_{\tvar}^i, q_{\tvar}^i \gets \text{Quadratize}\big(\marginfn^i(\cdot)\big)$ \textbf{for} $(\tvar, \marginfn^i) \in \pinch^i$,~\textbf{else}~$0$\;
    $R_\tvar^i,r^i_\tvar \gets\text{Quadratize}\big(\regularization \|\control_\tvar^i\|^2\big),\forall i,\tvar$\;
    $Z^i \gets Q^i_\horizon, z^i \gets q^i_\horizon$ \tcp*{terminal value functions}
    \For(\tcp*[f]{solve modified LQ game}){$\tvar = \horizon - 1, \dots, \tinit$}{
        $(\tilde \feedback^i_\tvar, \tilde \feedforward^i_\tvar) \gets$~coupled Riccati soln \cite[Cor. 6.1]{basar1998dynamic} \label{line:coupled-riccati}\;
        \For{$i = 1, \dots, \numplayers$}{
            \eIf(\tcp*[f]{critical time for P$i$}){$(\tvar, \cdot) \in \pinch^i$}{
                $Z^i_\tvar \gets Q^i_\tvar, z^i_\tvar \gets q^i_\tvar$ \tcp*{ignore future $\taux > \tvar$}
            }{
                $Z^i_\tvar, z^i_\tvar \gets$~std Riccati update \cite[Cor. 6.1]{basar1998dynamic}
            }
        }
    }
	\Return $(\tilde\bfeedback^i,\tilde\bfeedforward^i)_{i=1}^\numplayers$ 
\end{algorithm}

%% file: experiments.tex
\section{\revise{Application to Safe Autonomous Driving}}
\label{sec:application}

We showcase our approach in two case studies motivated by autonomous driving. The code is available online.\footnote{\looseness=-1 \href{https://github.com/SafeRoboticsLab/Reach-Avoid-Games}{\texttt{github.com/SafeRoboticsLab/Reach-Avoid-Games}}}

\subsection{\revise{Two-Player Defensive Driving}}

\begin{figure}
    \centering
    \begin{subfigure}[b]{\columnwidth}
        \centering
        \includegraphics[width=\linewidth]{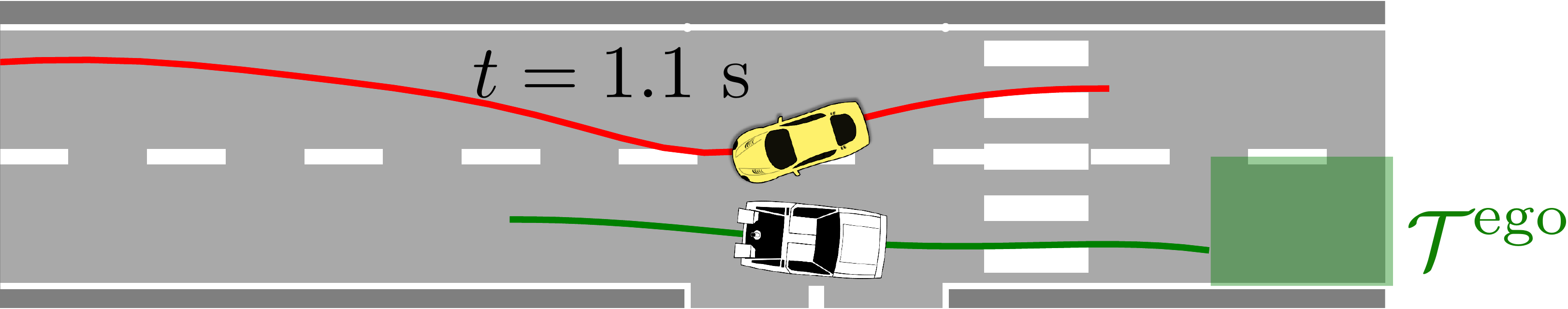}
        \caption{$\treact = $ 10 time steps $(\SI{1}{\second})$, collision successfully avoided}
    \end{subfigure}
    ~
    \par
    \begin{subfigure}[b]{\columnwidth}
        \centering
        \includegraphics[width=\linewidth]{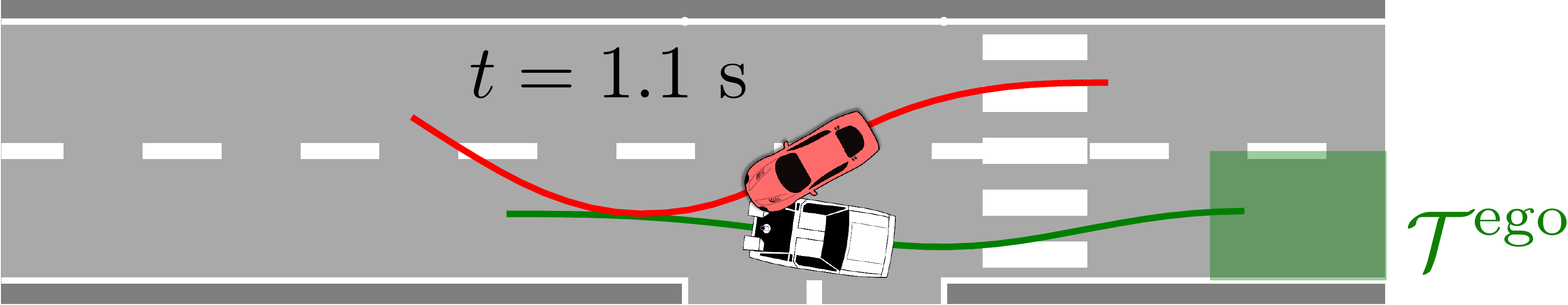}
        \caption{$\treact = $ 20 time steps $(\SI{2}{\second})$, collision unavoidable}
    \end{subfigure}
    \caption{Time-consistent reach-avoid method applied to a defensive driving problem, where the oncoming vehicle follows worst-case behavior until an intermediate time $\treact$. 
    For small $\treact$, the ego vehicle (white) avoids collision, but as $\treact$ increases collision becomes unavoidable.
    Oncoming vehicle color indicates adversarial (red) and cooperative (yellow) phase.}
    \label{fig:defensive-driving}
\end{figure}

\looseness=-1
Reach-avoid problems are widely relevant to a variety of domains.
To demonstrate the theoretical and algorithmic contributions of \cref{sec:methods}, we \revise{first} showcase them in a defensive driving scenario based upon that of \cite{chiu2021encoding}, illustrated in \cref{fig:defensive-driving}.
Here, we see an ``ego'' vehicle passing an oncoming car on a two-lane street.
The ego vehicle wishes to reach the end of its lane while avoiding collision with the oncoming car.
Following \cite{chiu2021encoding}, we encode a ``defensive driving'' mentality for the ego vehicle in a noncooperative dynamic game where, for some part of the horizon $\tvar \in \{\tinit, \dots, \treact - 1\}$ the oncoming vehicle wishes to collide, and thereafter its driver ``reacts'' and seeks to avoid collision.
We make several key changes to the formulation of \cite{chiu2021encoding}.
First, our formulation is a reach-avoid game, rather than a time-additive game. 
Second, we allow the ego player to choose both vehicles' control inputs after $\treact$. 
This allows the oncoming vehicle to behave truly adversarially \revise{until $\treact$}, and rewards it for causing an \revise{inevitable} collision at any point in the horizon, even after $\treact$.

We formulate the game as follows. Each vehicle \revise{is modeled as a kinematic bicycle \cite{sastry1991nonlinear} which presumes that tires do not slip. Here, the} state vector $\state^i := (p_x^i, p_y^i, \theta^i, \phi^i, v^i)$ \revise{is} comprised of \revise{center rear-axle} position, heading, front wheel angle, and speed.
The overall game state is then $\state = (\state^{\textnormal{ego}}, \state^{\textnormal{onc}})$ and each vehicle's states evolve as
\begin{multline}
    \label{eq:bicycle-dynamics}
    \frac{\state^i_{\tvar+1} - \state^i_\tvar}{\Delta t} = \Big(
        v^i \cos\theta^i,
        v^i \sin\theta^i,
        \frac{v^i \tan\phi^i}{L^i},
        \omega^i,
        a^i\Big)\,,
\end{multline}
\looseness=-1
where $\omega^i$ and $a^i$ are the front wheel rate and longitudinal acceleration of car $i$, $L^i$ is the \revise{wheelbase} ($\SI{4}{\meter}$ here) and $\Delta \tvar$ is the discrete time step ($\SI{0.1}{\second}$).
The ego's control input is then 
\begin{equation}
    \control^{\textnormal{ego}}_\tvar = \begin{cases}
        (\omega^{\textnormal{ego}}, a^{\textnormal{ego}}), &\tinit \le \tvar < \treact\\
        \big(\omega^{\{\textnormal{ego}, \textnormal{onc}\}}, a^{\{\textnormal{ego}, \textnormal{onc}\}}\big), &\treact < \tvar \le \horizon
    \end{cases}\,,
\end{equation}
and the oncoming vehicle's input affects only the first $\treact$ steps, i.e. $\control^{\textnormal{onc}}_\tvar = (\omega^{\textnormal{onc}}, a^{\textnormal{onc}})$ for only $\tinit \le \tvar < \treact$. 

The ego's objective is of the form \cref{eq:reach-avoid-objective}.
The failure \revise{set $\failureset^\text{ego}_\tvar$ contains all joint states in which (a) the ego vehicle and the oncoming vehicle are in collision or violating road boundaries, (b) the ego vehicle's steering angle exceeds a fixed range (here, $\pm30^\circ$) and (c) the oncoming vehicle violates its own analogous constraints for $\tvar>\treact$.}
\revise{The failure margin function $\failurefn^{\textnormal{ego}}_\tvar(\cdot)$ encodes signed distance to the above-defined set with the positive-inside convention.}
Similarly, the target function $\targetfn^{\textnormal{ego}}_\tvar(\cdot)$ encodes the ego vehicle's signed distance to
\revise{a goal region in the plane, illustrated in \cref{fig:defensive-driving}, in this case with the negative-inside convention.}

The oncoming car seeks to collide with the ego vehicle or force it to drive off the road \emph{before} the oncoming vehicle leaves the road.
Its target margin function $\targetfn^{\textnormal{onc}}_\tvar(\cdot)$ encodes the negative of the maximum between (a) the signed distance between the vehicles, and (b) the signed distance between the ego car and the road boundaries (with safety corresponding to positive values). 
The failure margin function $\failurefn^{\textnormal{onc}}_\tvar(\cdot)$ encodes signed distance to the road boundaries. 

We display the results of \cref{alg:main-alg} with the time-consistent reach-avoid \ac{lq} subroutine from \cref{alg:time-consistent-subroutine} in \cref{fig:defensive-driving} for several different values of $\treact$.
For low values of $\treact$, the ego vehicle has ample time to avoid the adversarial oncoming vehicle since adversarial control is only active over a short time interval.
At larger values of $\treact$, however, the ego vehicle cannot reach the target set without avoiding the failure set, and therefore fails to satisfy the reach-avoid condition of \cref{eq:reach-avoid-condition}.
By varying $\treact$, we are thus able to generate a range of increasingly strong defensive maneuvers.

\subsection{\revise{Three-Player Intersection}}

\begin{figure}
    \centering
    \includegraphics[width=\linewidth]{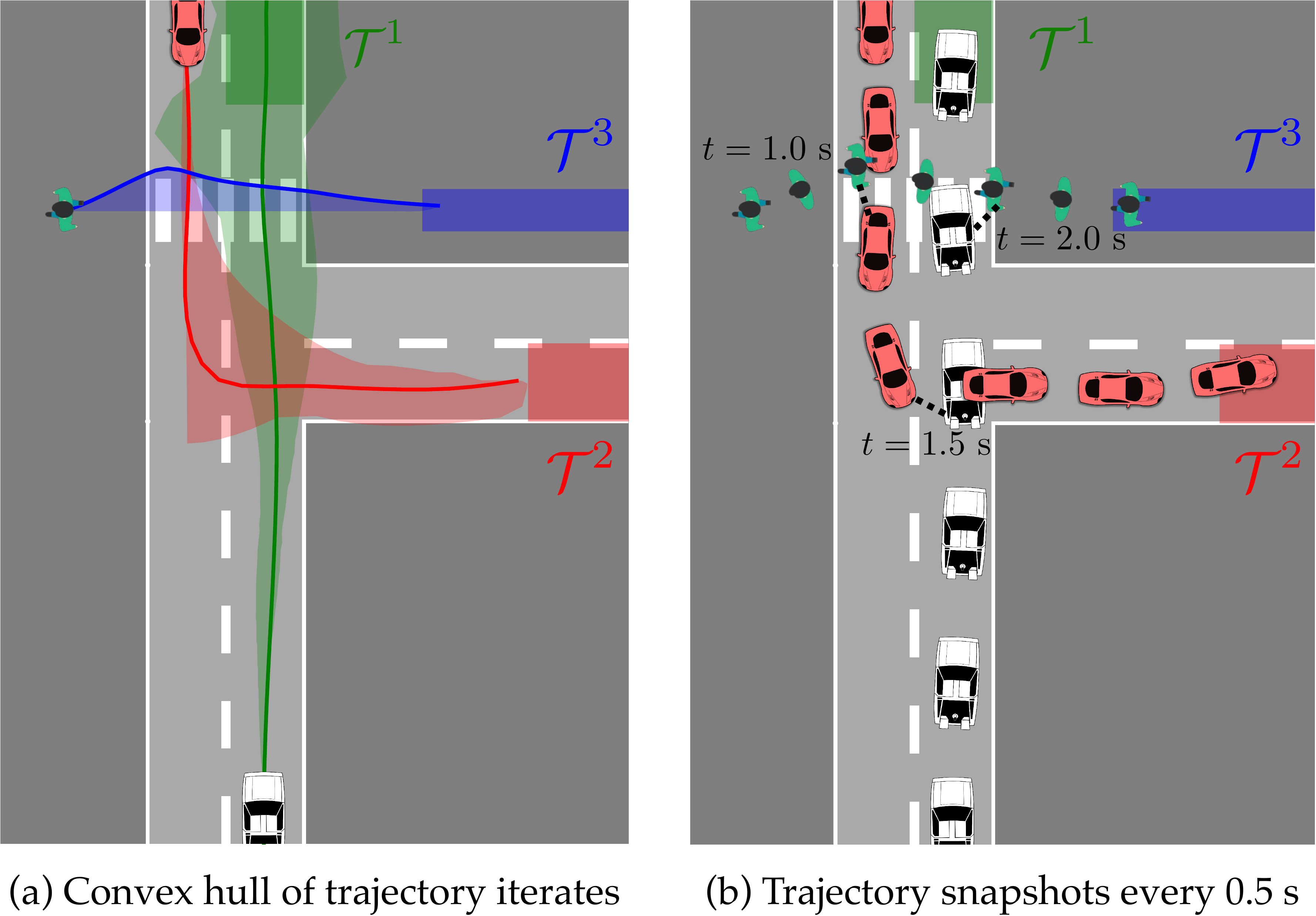}
    \caption{\revise{Two cars and a pedestrian negotiate a T-intersection.
    Each agent wishes to \emph{reach} a target region (respectively indicated as a colored rectangle) while \emph{avoiding} collision with others or, in the case of the two cars, transgressing the road boundaries. The solution computed by the proposed time-consistent reach-avoid method has all players reach their targets while avoiding all defined violations.
    Close approaches between agents are indicated by dashed lines and timestamped for reference.
    }
    }
    \label{fig:intersection}
\end{figure}

\revise{We extend the two-player driving scenario above by modeling one pedestrian and two cars who wish to cross a T-intersection modeled on that of \cite{fridovich2020efficient, fridovich2021approximate}.
As illustrated in \cref{fig:intersection}, one car wishes to drive straight through the intersection, while another wishes to make a left turn.
Both are modeled as in~\cref{eq:bicycle-dynamics}.
Meanwhile, a pedestrian with bounded instantaneous velocity control $(v_x^i, v_y^i)$ wishes to cross the intersection at a crosswalk.
Each agent has a reach-avoid objective of the form \cref{eq:reach-avoid-objective}, where the failure set contains collisions and, in the case of vehicles, leaving the allowed lanes, and the target set contains states with position greater than a desired distance ahead.
Our proposed time-consistent algorithm finds safe trajectories for all agents.
}


%% file: conclusion.tex
\section{Conclusion}
\label{sec:conclusion}


We study local algorithms for finding solutions to reach-avoid problems.
Our three core contributions are: (a) an efficient algorithm for reach-avoid problems which extends the reachability-only case of \cite{fridovich2021approximate}, (b) a time-consistent variation upon this algorithm whose output is robust to suboptimal or unanticipated behavior, and (c) derivations of our methods in the multi-player game-theoretic setting.
In particular, we prove both the time-\emph{in}consistency of the algorithm based upon \cite{fridovich2021approximate} and the (local) time consistency of our core algorithm.
We also demonstrate our approach in a simulated robust motion planning example which models a range of defensive driving behavior.

\looseness=-1
While we are motivated by the clear applications of reach-avoid problems to safety assurance in motion planning for autonomous vehicles, we also believe it is highly relevant to a broader range of robust control practitioners working in robotic grasping, surgical robotics, aerial collision avoidance, etc.
Future work will be investigate these applications and seek to validate the reliability and computational efficiency of our work in these varied physical systems.

%% file: references.bib
@article{kavraki1996probabilistic,
  title={Probabilistic roadmaps for path planning in high-dimensional configuration spaces},
  author={Kavraki, Lydia E and Svestka, Petr and Latombe, J-C and Overmars, Mark H},
  journal={IEEE Transactions on Robotics and Automation},
  volume={12},
  number={4},
  pages={566--580},
  year={1996},
  publisher={IEEE}
}

@book{borrelli2017predictive,
  title={Predictive control for linear and hybrid systems},
  author={Borrelli, Francesco and Bemporad, Alberto and Morari, Manfred},
  year={2017},
  publisher={Cambridge University Press}
}

@techreport{isaacs1954differential,
  title={Differential Games {I--IV}},
  author={Isaacs, Rufus},
  year={1954},
  institution={RAND Corporation},
  url={https://www.rand.org/pubs/research_memoranda/RM1391.html}
}

@book{sastry1991nonlinear,
  title={Nonlinear systems: analysis, stability, and control},
  author={Sastry, Shankar},
  year={1991},
  publisher={Springer}
}

@inproceedings{fridovich2020efficient,
  title={Efficient iterative linear-quadratic approximations for nonlinear multi-player general-sum differential games},
  author={Fridovich-Keil, David and Ratner, Ellis and Peters, Lasse and Dragan, Anca D and Tomlin, Claire J},
  booktitle={IEEE International Conference on Robotics and Automation (ICRA)},
  pages={1475--1481},
  year={2020},
  doi={10.1109/ICRA40945.2020.9197129}
}

@article{ratliff2016characterization,
  title={On the characterization of local {Nash} equilibria in continuous games},
  author={Ratliff, Lillian J and Burden, Samuel A and Sastry, S Shankar},
  journal={IEEE Transactions on Automatic Control},
  volume={61},
  number={8},
  pages={2301--2307},
  year={2016},
  publisher={IEEE}
}

@article{laine2021computation,
  title={The Computation of Approximate Generalized Feedback {Nash} Equilibria},
  author={Laine, Forrest and Fridovich-Keil, David and Chiu, Chih-Yuan and Tomlin, Claire},
  journal={arXiv preprint arXiv:2101.02900},
  year={2021},
  url={https://arxiv.org/abs/2101.02900}
}

@inproceedings{di2019newton,
  title={Newton’s method and differential dynamic programming for unconstrained nonlinear dynamic games},
  author={Di, Bolei and Lamperski, Andrew},
  booktitle={IEEE Conference on Decision and Control (CDC)},
  pages={4073--4078},
  year={2019},
  doi={10.1109/CDC40024.2019.9029237}
}

@book{basar1998dynamic,
  title={Dynamic noncooperative game theory},
  author={Ba{\c{s}}ar, Tamer and Olsder, Geert Jan},
  year={1998},
  publisher={SIAM}
}

@techreport{bellman1956dynamic,
  title={Dynamic programming},
  author={Bellman, Richard},
  year={1956},
  institution={RAND Corporation}
}

@inproceedings{van2014iterated,
  title={Iterated {LQR} smoothing for locally-optimal feedback control of systems with non-linear dynamics and non-quadratic cost},
  author={van den Berg, Jur},
  booktitle={IEEE American Control Conference (ACC)},
  pages={1912--1918},
  year={2014},
  url={https://ieeexplore.ieee.org/document/6859404}
}

@book{jacobson1970differential,
  title={Differential dynamic programming},
  author={Jacobson, David H and Mayne, David Q},
  year={1970},
  publisher={Elsevier}
}

@article{karaman2011sampling,
  title={Sampling-based algorithms for optimal motion planning},
  author={Karaman, Sertac and Frazzoli, Emilio},
  journal={The International Journal of Robotics Research (IJRR)},
  volume={30},
  number={7},
  pages={846--894},
  year={2011},
  publisher={Sage Publications Sage UK: London, England},
  doi={10.1177/0278364911406761}
}

@article{lavalle1998rapidly,
  title={Rapidly-exploring random trees: A new tool for path planning},
  author={LaValle, Steven M and others},
  year={1998},
  publisher={Ames, IA, USA},
  url={http://msl.cs.illinois.edu/~lavalle/papers/Lav98c.pdf}
}

@inproceedings{chiu2021encoding,
  title={Encoding Defensive Driving as a Dynamic {Nash} Game},
  author={Chiu*, Chih-Yuan and Fridovich-Keil*, David and Tomlin, Claire J},
  booktitle={IEEE International Conference on Robotics and Automation (ICRA)},
  year={2021},
  doi={10.1109/ICRA48506.2021.9560788}
}

@inproceedings{fridovich2021approximate,
  title={Approximate Solutions to a Class of Reachability Games},
  author={Fridovich-Keil, David and Tomlin, Claire J},
  booktitle={IEEE International Conference on Robotics and Automation (ICRA)},
  year={2021}
}

@book{Nocedal:2006,
	Address = {New York, NY, USA},
	Author = {Jorge Nocedal and Stephen J. Wright},
	Edition = {2nd Ed.},
	Publisher = {Springer},
	Title = {Numerical Optimization},
	Year = {2006}}

@article{alpern1985defining,
  title = {Defining Liveness},
  author = {Alpern, Bowen and Schneider, Fred B.},
  year = {1985},
  journal = {Information Processing Letters},
  volume = {21},
  number = {4},
  pages = {181--185},
  issn = {00200190},
  doi = {10.1016/0020-0190(85)90056-0}
}

@article{evans1984differential,
  title = {Differential Games and Representation Formulas for Solutions of {{Hamilton}}-{{Jacobi}}-{{Isaacs}} Equations},
  author = {Evans, L. C. and Souganidis, P. E.},
  year = {1984},
  journal = {Indiana University Mathematics Journal},
  volume = {33},
  number = {5},
  pages = {773--797},
  issn = {0022-2518},
  url = {https://www.jstor.org/stable/45010271}
}

@inproceedings{fisac2015reachavoid,
  title = {Reach-Avoid Problems with Time-Varying Dynamics, Targets and Constraints},
  booktitle = { {{ACM International Conference}} on {{Hybrid Systems}}: {{Computation}} and {{Control}} {(HSCC)}},
  author = {Fisac, Jaime F. and Chen, Mo and Tomlin, Claire J. and Sastry, S. Shankar},
  year = {2015},
  pages = {11--20},
  doi = {10.1145/2728606.2728612},
  isbn = {978-1-4503-3433-4}
}

@article{bokanowski2010reachability,
  title = {Reachability and Minimal Times for State Constrained Nonlinear Problems without Any Controllability Assumption},
  author = {Bokanowski, Olivier and Forcadel, Nicolas and Zidani, Hasnaa},
  year = {2010},
  journal = {SIAM Journal on Control and Optimization},
  volume = {48},
  number = {7},
  pages = {4292--4316},
  publisher = {{Society for Industrial and Applied Mathematics}},
  issn = {0363-0129},
  doi = {10.1137/090762075}
}

@article{mitchell2008flexible,
  title = {The Flexible, Extensible and Efficient {Toolbox} of~{{Level~Set Methods}}},
  author = {Mitchell, Ian M.},
  year = {2008},
  journal = {Journal of Scientific Computing},
  volume = {35},
  number = {2},
  pages = {300--329},
  issn = {1573-7691},
  doi = {10.1007/s10915-007-9174-4}
}
